%% file: main.tex
\documentclass[%
  twocolumn,
superscriptaddress,
 amsmath,amssymb,
 aps, prl,
]{revtex4-2}
\usepackage{tabularx}
\usepackage{ltablex}
\usepackage{multirow}
\usepackage{amssymb}
\usepackage{braket}
\usepackage[dvipsnames]{xcolor}
\usepackage{xcolor}
\usepackage{graphicx}
\usepackage{dcolumn}
\usepackage{bm}

\usepackage[colorlinks=true,
            linkcolor=blue,
	          citecolor=blue,
	          urlcolor=blue,
            hypertexnames=false
            ]{hyperref}

\newcommand{\trieste}{Dipartimento di Fisica, Universit\`a di Trieste,  I-34151 Trieste, Italy}
\newcommand{\sissa}{Scuola Internazionale Superiore di Studi Avanzati (SISSA), I-34136 Trieste, Italy}        
\begin{document}
\title{Fluctuations and Correlations of Local Topological Order Parameters\\in Disordered Two-dimensional Topological Insulators}%

\author{Roberta Favata}
\affiliation{\trieste}
\email{roberta.favata@phd.units.it}
\author{Nicolas Baù}
\affiliation{\trieste}
\author{Antimo Marrazzo}
\affiliation{\sissa}
\email{amarrazz@sissa.it}

\date{\today}

\begin{abstract}
Two-dimensional topological insulators are characterized by an insulating bulk and conductive edge states protected by the nontrivial topology of the bulk electronic structure. They remain robust against moderate disorder until Anderson localization occurs and destroys the topological phase. Interestingly, disorder can also induce a topological phase\textemdash known as a topological Anderson insulator\textemdash starting from an otherwise pristine trivial phase. While topological invariants are generally regarded as global quantities, we argue that space-resolved topological markers can act as local order parameters, revealing the role of fluctuations and correlations in the local topology under Anderson disorder and vacancies. With this perspective, we perform numerical simulations of disorder-driven topological phase transitions in the Haldane and Kane-Mele models, using supercells with both open and periodic boundary conditions. We find that short-scale fluctuations of topological markers vanish upon coarse-graining, except at the topological phase transition, where their correlation length peaks and large-scale fluctuations remain. Notably, such a topological correlation function is characterized by critical exponents that appear universal across disorder types, yet they can resolve different topological phase transitions.
\end{abstract}

\maketitle


Topological insulators are often discussed as phases of matter that are not characterized by a local order parameter, hence falling outside Landau's paradigm based on symmetry breaking~\cite{bernevig_book_2013}. In particular, topological invariants are usually defined as global properties of the ground-state wavefunction in reciprocal space: a fundamental example is the Chern number, which can be calculated as the integral of the Berry curvature over the entire Brillouin zone (BZ) and is an integer~\cite{thouless_prl_1982,Vanderbilt2018}. The situation is similar even in a more general many-body framework, as the Chern number admits a similar expression where the Bloch wavevectors are replaced by the phase parameters in the boundary conditions~\cite{niu_prb_1985}: that makes it a global property of the full many-body wavefunction on the torus. In this sense, topological invariants are non-local properties that do not depend on the local details of the system but describe global features that are robust to adiabatic transformations. Their global character is a description of the wavefunction geometry with respect to a closed parameter space, oftentimes the BZ torus. For the sake of simplicity, from now on we restrict the discussion to non-interacting systems in a ``mean-field'' sense, where the Chern number and other topological invariants are defined in reciprocal space, although our conclusions might be more general. In this case, the non-local character stems from a definition that assumes to deal with systems described by the Bloch theorem, essentially crystals with periodic boundary conditions (PBCs). 

However, the use of PBCs is no more than a convenient artifact and real systems are always finite: so it should be (and it is) possible to define topological phases in real space. In addition, the locality (or “nearsightedness” according to Kohn~\cite{Kohn1996}) of the ground-state electronic structure demands not only a real-space description, but a \emph{local} formulation, i.e., solely based on the knowledge of the ground-state in the neighborhood around the region of interest. Beyond the conceptual motivation, more pragmatic applications loom larger: real materials are never perfect and can be affected by disorder such as defects, while non-crystalline systems\textemdash including quasi-crystals \cite{huang_prl_2018,huang_prb_2018} and amorphous materials~\cite{fazzio_nanolett_2019,marsal_pnas_2020,corbae_natmat_2023,corbae_EPL_2023}\textemdash are known to exhibit non-trivial topological phases.

In this work, we argue that local topological markers~\cite{bianco_prb_2011,marrazzo_prb_2017,pbc_lcm,z2_marker} not only allow probing topological phases in non-crystalline materials, but they play the role of \emph{bona fide} local order parameters.  We discuss numerical results for topological phase transitions driven by disorder or vacancies, where local topological markers are shown to exhibit meaningful real-space fluctuations and two-point correlations. We demonstrate how the scale of fluctuations and the Chern-Chern correlation length increase as the topological phase transition is approached, which resonates with conventional (non-topological) phase transitions. A proper macroscopic average of the local markers averages out the small-scale fluctuations and returns an integer topological invariant in the thermodynamic limit, essentially performing a coarse-graining procedure similar to decimation~\cite{kadanoff_1966,wilson_prl_1972,kadanoff_prb_1975, Wilson1979,wilson_rmp_1983}. While topological invariants describe robust properties of the wavefunction, topological phases admit a description in terms of local order parameters and of their small-scale fluctuations and correlations, providing insights into the physics of disorder-driven topological phase transitions.

Over the years, several approaches have been proposed to calculate topological invariants in non-crystalline systems, a non-exhaustive list includes the switch-function formalism~\cite{Elbau2002, PhysRevB.97.195312, 10.1063/5.0096720}, methods based on the scattering matrix~\cite{PhysRevB.85.165409, PhysRevB.83.155429} or on the non-commutative index theorem~\cite{Avron1994, AVRON1994220, 10.1063/1.5026964}, the structural spillage~\cite{segovia_prr_2023}, the Bott index~\cite{10.1063/1.3274817, HASTINGS20111699, Loring_2011}, real-space formulas~\cite{prodan_prl_2010,prodan_NJP_2010}, single-point formulas~\cite{ceresoli_sp_prb_2007,Favata_2023} and local markers~\cite{bianco_prb_2011,pbc_lcm,z2_marker}; here, we focus and briefly outline the last two approaches, which can tackle respectively  homogeneous and inhomogeneous systems. A popular way to approach the non-crystalline case is to consider large periodic cells such that the sampling of the BZ can be reduced to a single point in reciprocal space, usually the $\Gamma$ point. In the limit of large supercells, the Chern number can be calculated through single-point formulas as discussed in Ref.~\cite{ceresoli_sp_prb_2007} for the Chern number and in Ref.~\cite{Favata_2023} for the $\mathbb{Z}_2$ invariant. Single-point topological numbers are an effective way to probe the topology for non-crystalline systems, including crystals at finite temperature through molecular dynamics, but they lack spatial resolution. Instead, topological markers can probe the topology locally in real space and describe inhomogeneous systems such as topological-trivial superlattices or heterojunctions. Originally, local topological markers were formulated and validated in open boundary conditions (OBCs)~\cite{bianco_prb_2011,marrazzo_prb_2017}, where the Chern marker takes the form
\begin{equation}
  \label{eq:obc_lcm}
  \mathcal{C}(\mathbf r)= - 4\pi \text{Im}\bra{\mathbf r} \mathcal Px(\mathbb I-\mathcal P)y \ket{\mathbf r},
\end{equation}
with $\mathcal{P}$ the ground-state projector and $x,y$ the position operators in OBCs. More recently~\cite{pbc_lcm}, the Chern marker has been formulated also in PBCs, where it can take the form~\footnote{In the numerical simulations presented in this work, we used the symmetric version of Eq.~\ref{eq:pbclcm_3} $
	C^{(sym)}(\mathbf r)=-\frac{1}{8\pi}\mathrm{Im}\bra{\mathbf{r}} \left(\left[\mathcal{P}_{\mathbf{b}_1},\mathcal{P}_{\mathbf{b}_2}\right]+\left[\mathcal{P}_{-\mathbf{b}_1},\mathcal{P}_{-\mathbf{b}_2}\right] \right.   - \left.\left[\mathcal{P}_{-\mathbf{b}_1},\mathcal{P}_{\mathbf{b}_2}\right]-\left[\mathcal{P}_{\mathbf{b}_1},\mathcal{P}_{-\mathbf{b}_2}\right]\right)\mathcal{P}_{\Gamma} \ket{\mathbf{r}}$   that converges faster than~Eq.~\ref{eq:pbclcm_3} (asymmetric formula), as discussed in Ref.~\cite{pbc_lcm}.}
\begin{eqnarray}
  \label{eq:pbclcm_3}
	\mathcal{C}(\mathbf r)=-\frac{1}{2\pi}\mathrm{Im}\braket{\mathbf{r}|\big[ \mathcal P_{\mathbf b_1}, \mathcal P_{\mathbf b_2} \big] \mathcal P_{\Gamma} |\mathbf{r}}.
\end{eqnarray}
with $\mathcal P_{\mathbf b_j}=\sum_{n=1}^{N_{occ}}\ket{\tilde u_{n\mathbf b_j}}\bra{\tilde u_{n\mathbf b_j}}$ being the ground-state projectors at BZ edge in the single-point limit. The projectors $\mathcal P_{\mathbf b_j}$, where $\mathbf{b}_{j}$ are the reciprocal lattice vectors, are expressed in terms of the ``dual'' states $|\tilde u_{n\mathbf b_j}\rangle$ and represent, in the limit of a large supercell, the states obtained by parallel transport:
\begin{equation}
  \label{eq:duals_def}
  \ket{\tilde u_{n\mathbf b_j}}=\sum_{m=1}^{N_{occ}}S^{-1}_{mn}(\mathbf b_j)e^{-i\mathbf b_j\cdot\mathbf r}\ket{u_{m\Gamma}},
\end{equation}
where we make use of the overlap matrix $S_{nm}(\mathbf b_j) = \braket{u_{n\Gamma}|e^{-i\mathbf b_j\cdot\mathbf r}|u_{m\Gamma}}$ as discussed in Refs.~\cite{ceresoli_sp_prb_2007,Favata_2023}.
The local Chern number can then be obtained by calculating the macroscopic average of the Chern markers, either in OBCs (Eq.~\ref{eq:obc_lcm}) or PBCs (Eq.~\ref{eq:pbclcm_3}), essentially taking the trace per unit of area ($\mathrm{Tr}_A$) in a neighborhood around the region of interest. Notably, the trace of the PBC Chern marker over the entire supercell is equal to the single-point Chern number of Ceresoli and Resta~\cite{ceresoli_sp_prb_2007}, as shown in Ref.~\cite{pbc_lcm}.
Hence, we can study the effect of disorder on topological phases from a more global perspective, by averaging the single-point Chern number $C_{sp}$ over multiple disorder realizations, and from a local viewpoint through real-space maps of macroscopic averages of the OBC or PBC Chern markers $\mathcal{C}(\mathbf r)$.
\begin{figure*}[t]
  \includegraphics[width=\textwidth]{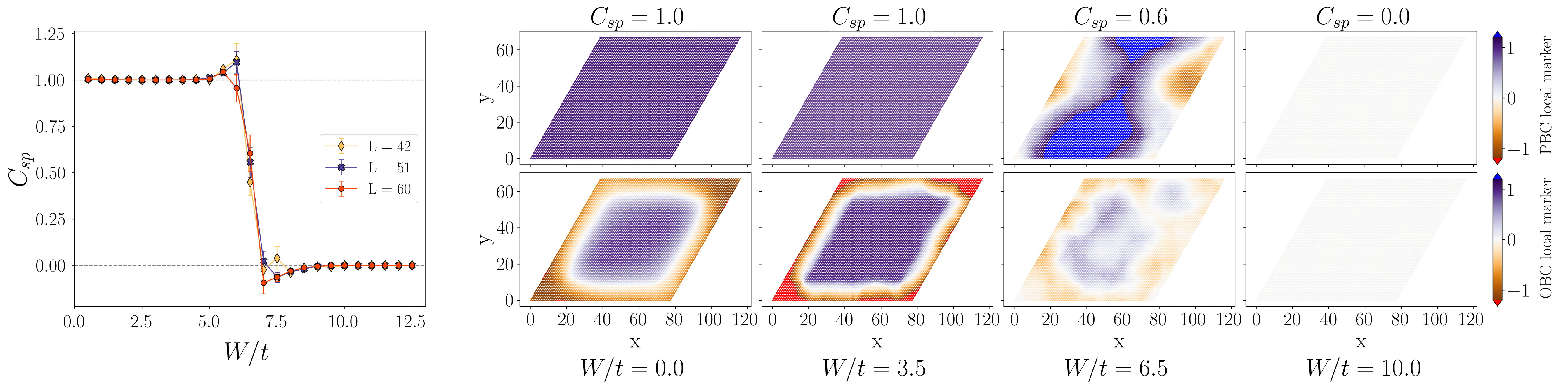}
  \caption{\label{fig:topo-to-triv}Topological-to-trivial phase transition driven by Anderson disorder in the Haldane model ($\Delta/t=1.25$). Left panel: single-point Chern number ($C_{sp}$) calculations as a function of the disorder strength $W$ [in units of the nearest-neighbor hopping $t$]. Right panel: PBC (top row) and OBC (bottom row) Chern marker maps for the linear size $L=78$ ($12168$ sites) for various values of $W$; the single-point Chern number corresponding to the maps is displayed above. The disordered on-site terms are the same in PBCs and OBCs. The color (gray) scale displays values in $[-1.2,1.2]$, values outside the range are marked in red (lower than -1.2) and blue (larger than 1.2). The presence of disorder induces small-scale fluctuations that are averaged out even for strong disorder amplitudes by coarse graining (a.k.a. macroscopic average), i.e., by taking the trace over the neighborhood sites within a radial cutoff $r_c=20$. Fluctuations increase close to the phase transition, where finite-size effects are sensibly stronger for finite samples. }
\end{figure*}

Here, we consider the Haldane model~\cite{haldane_1988} on the honeycomb lattice in the presence of Anderson disorder~\cite{abrahams_prl_1979}; numerical simulations for the Kane-Mele model~\cite{kane_quantum_2005,kane_z2_2005} are reported later and in the Supplementary Material (SM)~\cite{SM}. For the pristine case, the nearest-neighbor hopping is real and set to $t=-4$, while the second-nearest neighbor hopping is complex with amplitude $|t_2/t| = 1/4$ and phase $\phi = -\frac{\pi}{2}$.
Disorder is introduced through a random on-site term with uniform distribution $w_i \in \left[ -\frac{W}{2},\frac{W}{2} \right]$, where $W$ is the disorder strength.
Single-point invariants and local markers are calculated through the \texttt{StraWBerryPy} Python package~\cite{strawberrypy}, which is part of the Wannier function software ecosystem~\cite{marrazzo_rmp_2024} and is interfaced to popular tight-binding engines such as \texttt{TBmodels}~\cite{tbmodels,TB} and \texttt{PythTB}~\cite{pythtb}.
In Figure~\ref{fig:topo-to-triv}, we show results of numerical simulations for the Haldane model as a function of $W$, starting from a pristine system with Chern number $C=1$ and on-site term $\Delta/t=1.25$: for weak disorder the system is topological, while it becomes trivial ($C=0$) for larger values, with a rather sharp transition in between. The left-hand panel of Fig.~\ref{fig:topo-to-triv} shows the single-point Chern number calculated for large supercells containing, respectively, 3528 ($L=42$), 5202 ($L=51$) and 7200 ($L=60$) sites; the points and error bars correspond to the average and standard deviation over 300 realizations. 
While for sufficiently strong disorder amplitudes the density of states becomes gapless, we confirm that the two phases are insulating by calculating the localization length~\cite{RS_prl_1999} $\lambda$, which remains finite and increases near the critical disorder strength (where it is expected that $\lambda$ diverges with the system size and the critical phase is metallic).
In the right-hand panel of Fig.~\ref{fig:topo-to-triv}, we show the real-space maps of the local Chern number, both in PBCs and OBCs for $L=78$ ($12168$ sites), where the value at each site is calculated by averaging Chern markers over all the primitive cells within a radial cutoff $r_c=20$ (in units of the lattice parameter). Remarkably, even if the disorder amplitude increases to values comparable with the on-site term of the pristine case, the local topology depicts a homogeneous system throughout the transition, except close to the critical point where the single-point Chern number deviates from integer values and real-space fluctuations are strong. In that region, finite-size effects are more visible especially for finite samples: the OBC simulations return rather different topological maps with respect to their PBC counterparts, even if all the on-site terms are equal and only the boundary conditions differ.

\begin{figure*}[t]
  \includegraphics[width=\textwidth]{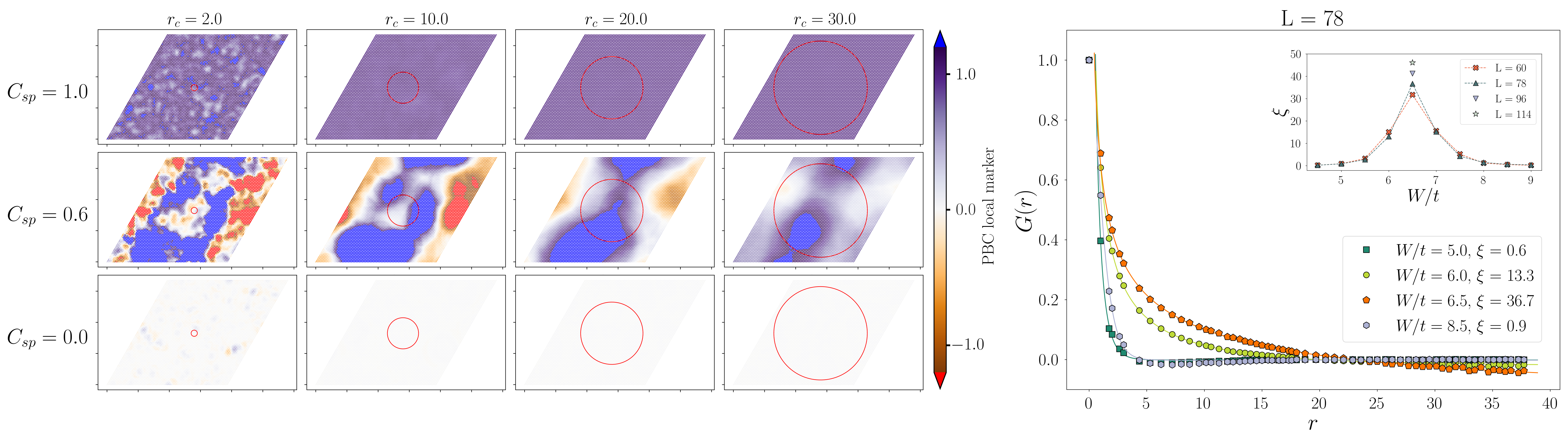}
  \caption{Macroscopic average and Chern-Chern correlation function. Left panel: PBC Chern marker maps for three different values of the disorder strength $W/t$, which correspond to three different values of the single-point Chern number ($C_{sp}=1.0$ for $W/t = 5$, $C_{sp}=0.6$ for $W/t = 6.5$ and $C_{sp}=0.0$ for $W/t = 10$). Maps are drawn for increasing values of the cutoff $r_c$ over which the macroscopic average is performed. The cutoff $r_c$ is marked by the radius of the red circle centered in the middle of the sample. Short-scale fluctuations of the topological marker average out in the topological and trivial phase, while large-scale fluctuations remain at the transition. Right panel: the connected Chern-Chern radial correlation function (Eq.~\ref{eq:corr}) computed at different disorder strengths, averaged over $200$ realizations with linear size $L = 78$. The correlation length $\xi$ is obtained by fitting with the Ornstein-Zernike formula (Eq.~\ref{eq:oz}); the inset reports $\xi$ as a function of $W/t$ for system sizes $L=60$ and $L=78$, additional values for $L=96,114$ are shown at $W/t = 6.5$ only. \label{fig:fluc_corr}}
\end{figure*}

Next, we investigate the length scale of the fluctuations of local topological markers.
The left-hand panel of Figure~\ref{fig:fluc_corr} shows real-space maps for the Chern number in PBCs computed for increasing cutoff radii (red circles) ranging from $r_c=2$ to $r_c=30$, in the topological ($C_{sp}=1.0$, $W/t = 5$) and trivial ($C_{sp}=0.0$, $W/t = 10$) phase, as well as close to the topological phase transition ($C_{sp}=0.6$, $W/t = 6.5$).
If the trace is taken over a few sites, the local Chern marker displays fluctuations in all phases, that become very strong close to the phase transition. If we average over larger regions, essentially coarse graining the topological markers, short-scale fluctuations disappear both in the topological and trivial phase: already at $r_c=10$ these disordered systems appear topologically uniform. On the contrary, sufficiently close to the phase transition we observe large-scale fluctuations that persist even if the macroscopic average is performed over rather large regions (e.g., $r_c=30$).

We rationalize these results by studying the real-space correlations between local Chern numbers. 
We calculate the connected radial Chern-Chern correlation function as
\begin{equation}
  \label{eq:corr}
    G (r) = \braket{ \mathcal{C}_0  \mathcal{C}_r}_{\mathrm{samples}} -\braket{\mathcal{C}_0}^{2}_{\mathrm{samples}},
\end{equation}
where the mean is taken over $N_{samples}$ disorder realizations. The one- and two-point correlations are normalized as
\begin{equation}
\braket{ \mathcal{C}_0  \mathcal{C}_r}_{\mathrm{samples}} = \biggl\langle  \frac{1}{N_r}  \frac{ \sum_{i,j}  \delta(r - r_{ij}) \mathcal{C}_i^s \mathcal{C}_j^s } {\sigma^2_s} \biggr\rangle_{\mathrm{samples}},
\end{equation}
with $N_r$ being the number of sites $j$ at distance $r$ from site $i$, $\mathcal{C}_{i}^s$ are the local Chern markers at site $i$ in the realization $s$, $\langle \mathcal{C}_i\rangle_{s} $ is the mean value of the local Chern number over the system  (i.e., the single-point Chern number) and $\sigma^2_s = \langle \mathcal{C}_i^2 \rangle_{s} - \langle \mathcal{C}_i \rangle_{s}^2 $ is the corresponding variance. 
In the right-hand panel of Fig.~\ref{fig:fluc_corr}, we report the radial correlation function for local Chern markers computed at various disorder strengths, including those corresponding to the Chern maps in the left-hand panel, by averaging over $200$ realizations. The inset shows the correlation length $\xi$ extracted by fitting $G(r)$ with the Ornstein-Zernike form~\cite{fisher_jmp_1964,tuckerman_book_2023}:
\begin{equation}
  \label{eq:oz}
G(r) \sim r^{-p} e^{-r/\xi},
\end{equation}
where $p \simeq 1/2$ is obtained by regression close to the phase transition.
In both the topological and trivial phases the correlation length is in general very short, comparable with the pristine lattice parameter, no matter the large $W$ that might be present: this corresponds to the situation where the local Chern markers are mostly $1$ or $0$, fluctuations are small and short-scale. As we approach the phase transition, $G(r)$ acquires longer tails and the correlation length increases dramatically. We note that the large values ($\xi\approx 30-50$) obtained very close to the transition might not be exact estimates, precisely because the lengths are comparable with the simulation cell ($L=78$): in the inset, we show results up to $L=114$ (25992 sites), where $\xi$ remains lower than half of the cell length. It is worth noting that, while the local topological order parameter sharply distinguishes the two phases, its fluctuations and correlations display the same behavior on both sides of the transition and are characterized by the same exponent $p$ and length $\xi$ (see the inset of Fig.~\ref{fig:fluc_corr}).

These results highlight that local Chern markers behave as local order parameters describing disorder-driven phase transitions: fluctuations average out in the thermodynamic limit and occur at short scales, except at the boundary where the correlation length becomes rather long and large-scale fluctuations remain. Hence, the real-space fluctuations and correlations of local Chern markers are excellent tools to diagnose and discuss topological phase transitions in non-crystalline systems, where the original definition of the Chern number is of no avail. 

In the Appendices, we show that our conclusions also hold for topological Anderson insulators (TAIs)~\cite{TAI,TAI_Beenakker,TAI_phenomena}, where a pristine trivial phase, sufficiently close to the phase boundary, becomes topological due to small amounts of disorder before strong disorder eventually drives the system back to a trivial insulator via Anderson localization. Therein, we show that the same analysis applies also in the presence of lattice vacancies, where a topological-to-trivial transition is triggered once a critical density of random sites is removed.

Although we discussed Chern numbers and quantum anomalous Hall insulators (QAHIs) in the Haldane model, our framework is equally applicable to quantum spin Hall insulators (QSHIs). In the SM~\cite{SM}, we show real-space maps of the local topology for the Anderson-disordered Kane-Mele model~\cite{kane_quantum_2005,kane_z2_2005}, where the spin-Chern marker~\cite{z2_marker} behaves as a local $\mathbb{Z}_2$ order parameter.

\begin{figure}
  \includegraphics[width=\linewidth]{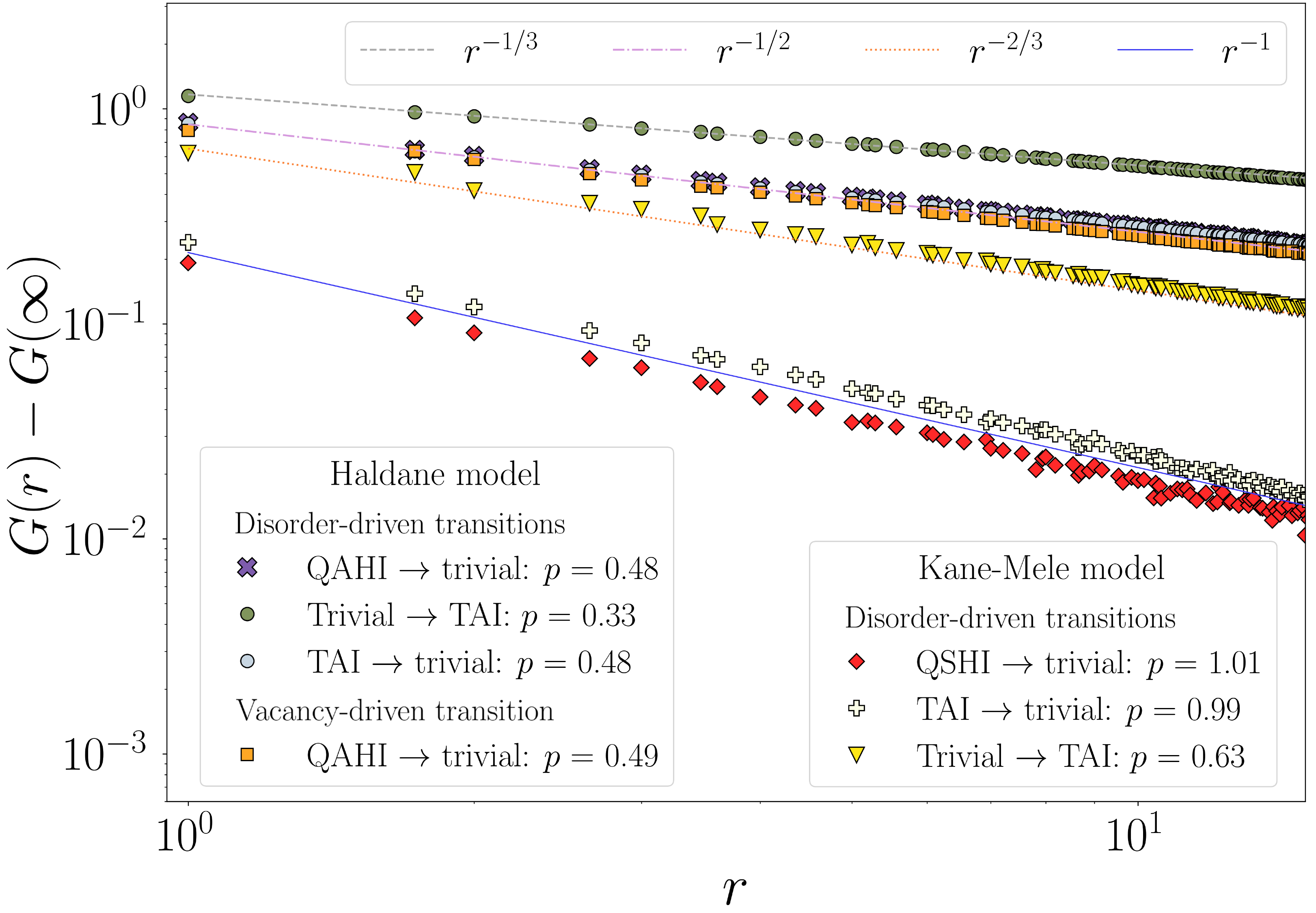}
  \caption{Power-law decay of the correlation function $G(r)$ for several disorder-driven topological phase transitions in the Haldane and Kane-Mele models. The dashed lines represent power laws $r^{-p}$ for rational values of $p$. All topological$\rightarrow$trivial transitions are characterized by exponents that do not depend on the type of disorder but only on the symmetry class: $p$ is about $1/2$ in QAHIs and $1$ in QSHIs. The trivial$\rightarrow$TAI transition is characterized by a different exponent, about $1/3$ in QAHIs and $2/3$ in QSHIs. The parameters of the Haldane model ($L=96$, except for the case of vacancies where $L=57$) are reported in the previous figures, and in the Appendices for the transitions from/to the TAI phase and in the presence of vacancies; here we list the values for the Kane-Mele case ($L=72$): the QSHI$\rightarrow$trivial transition occurs at about $W/t = 4$ with $\Delta/\lambda_{SOC} = 3$; the trivial$\rightarrow$TAI and the TAI$\rightarrow$trivial occur respectively at about $W/t = 1.4$ and $W/t = 4$ with $\Delta/\lambda_{SOC} = 5.3$. \label{fig:critical}}
\end{figure}

Such generality in our findings suggests that $p$ could serve as a universal critical exponent, independent of the specific choice of parameters. We report in Fig.~\ref{fig:critical} a comparison of the power-law decay of $G(r)$ and its exponent $p$ extracted from the data, for several disorder-driven transitions in both the Haldane and Kane-Mele models (the SM~\cite{SM} provides a detailed discussion of how finite-size effects are accounted for). Remarkably, all topological-to-trivial transitions are characterized by the same exponent, which does not depend on the type of disorder (i.e., Anderson or vacancies) but depends on the symmetry class: it differs between QAHIs and QSHIs. Notably, the trivial-to-TAI transition is characterized by a different exponent with respect to the topological-to-trivial cases, hence providing a fingerprint to distinguish TAI from disordered topological insulators. The ratio between the exponents in QSHIs and QAHIs is around 2 and does not depend on the type of transition: in QAHIs (QSHIs), $p$ is, respectively, about $1/2$ ($1$) for the topological-to-trivial and $1/3$ ($2/3$) for the trivial-to-TAI transitions.

Finally, we emphasize that fluctuations and correlations are particularly insightful whenever the topological phases lack translational invariance in the strict sense. Indeed, for clean samples there are no fluctuations owing to crystalline order, which forces the local topological markers to be uniform over all sample: any real-space fluctuation would break translational invariance. Hence, in the limit of perfectly periodic crystals, local topological order parameters are strictly quantized with vanishing fluctuations, in agreement with the standard definition of topological invariants on the BZ torus. 

While in this work we study topological phase transition at zero temperature and driven by disorder, the local Chern marker can be calculated also for metallic systems and in the presence of an electronic temperature, where it is non-quantized and represents the intrinsic geometric part of the local anomalous Hall conductivity~\cite{marrazzo_prb_2017,rauch_prb_2018}. As an outlook, it could be interesting to study temperature-driven topological phase transitions, such as those occurring in ZrTe$_5$~\cite{xu_prl_2018} and transition-metal dichalcogenides~\cite{khaustov_2dmat_2025}, by studying fluctuations and correlations of local topological order parameters. 

The data that support the findings of this article are openly available~\cite{data}.

The authors acknowledge CINECA, under the ISCRA initiative and the CINECA-UniTS and CINECA-SISSA agreements, for the availability of high-performance computing resources and support, including simulation time on Galileo100.

\bibliography{biblio}

\input{end_matter.tex}
\onecolumngrid
\section{Supplementary material}
\input{sm_text.tex}
\end{document}

%% file: end_matter.tex
\renewcommand\thesubsection{Appendix \arabic{subsection}}

\renewcommand\theequation{EM\arabic{equation}}
\appendix
\onecolumngrid
\section{Appendices}
\setlength{\parskip}{0pt}
\twocolumngrid
\begin{figure*}[t!]
  \includegraphics[width=\textwidth]{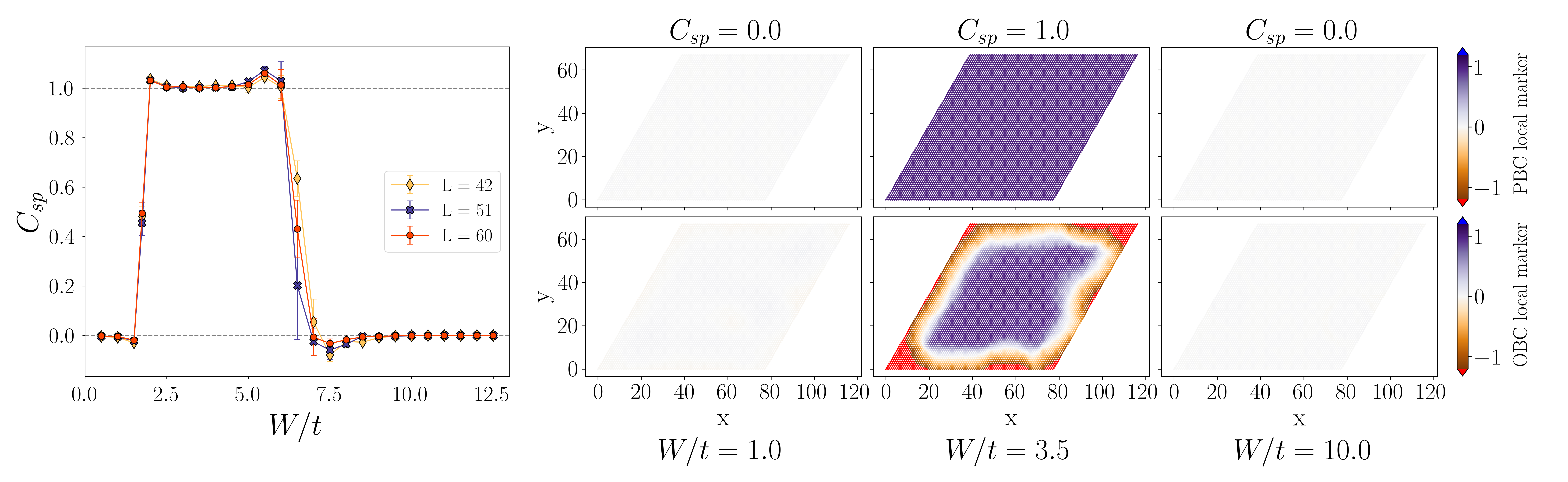}
  \caption{\label{fig:tai}Trivial-topological-trivial phase transition driven by Anderson disorder in the Haldane model ($\Delta/t=1.375$); the disorder-induced phase with $C_{sp}=1.0$ is the topological Anderson insulator. Plots are arranged as in Fig.~\ref{fig:topo-to-triv}. The local topology is homogeneous for all phases, independently of the disorder strength, except at the two phase transitions where the fluctuations become large, as signaled by the large variance of the single-point Chern numbers. }
\end{figure*}
\begin{figure*}[t!]
  \includegraphics[width=\textwidth]{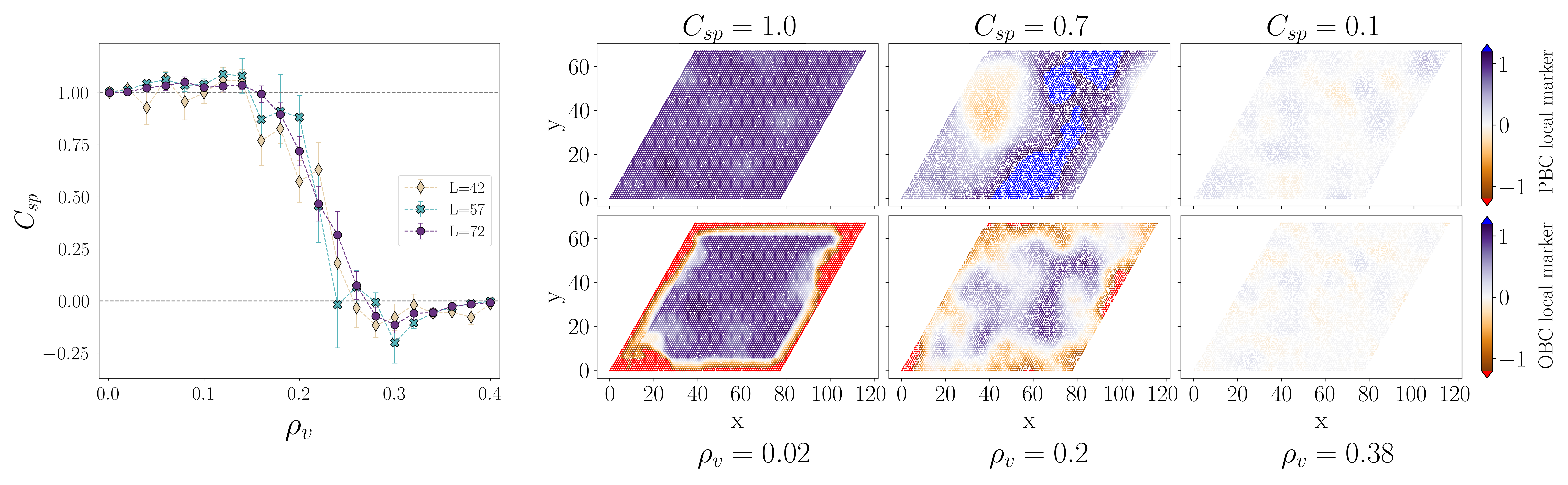}
  \caption{Topological phase transition in the Haldane model ($\Delta=0$) driven by the presence of vacancies, with relative density $\rho_v$. Plots are arranged as in Figs.~\ref{fig:topo-to-triv} and~\ref{fig:tai}. The macroscopic average is performed within a radial cutoff $r_c = 6$ and yields a rather uniform topological landscape, except in the neighborhood of the phase transition where it becomes patchy and strongly affected by finite-size effects. \label{fig:vacancies}}
\end{figure*}
\paragraph{Disorder-driven phase transitions in the TAIs\textemdash}TAIs exhibit two phase transitions: for increasing disorder, the first transition goes from the trivial phase to the TAI, where the topological invariant is non-trivial, which is followed by a second transition from the TAI to a trivial insulator owing to Anderson localization. In Fig.~\ref{fig:tai}, we report numerical simulations for phase transitions from and to the TAI phase in the Haldane model, starting from a pristine system with Chern number $C=0$ and on-site term $\Delta/t=1.375$ (all the other parameters are equal to those listed in the main text). Here we display both the single-point Chern number and PBC/OBC Chern marker maps (with $r_c = 10$), adopting the framework of Fig.~\ref{fig:topo-to-triv}. 
Also in this case, if the macroscopic average of the Chern markers is performed on the appropriate length scale\textemdash set by the Chern-Chern correlation length\textemdash the local topology is homogeneous for all phases, independently of the disorder strength. At both phase transitions, the fluctuations become large: this is signaled by the large variance over disorder realizations of the single-point Chern numbers and is related (through the fluctuation-dissipation relation) to the correlation length becoming larger. 

\paragraph{Vacancy-driven phase transition\textemdash}Now we consider a different type of disorder: a relative density of vacancies $\rho_v$ is introduced by randomly removing sites from the lattice. In Fig.~\ref{fig:vacancies} we report numerical simulations for the Haldane model as a function of $\rho_v$. The pristine model is in the topological phase ($\Delta = 0$, $\phi = -\frac{\pi}{8}$), whereas for large $\rho_v$ the Chern number becomes trivial; the transition occurs for about $20\%$ of removed sites. The left-hand panel displays the single-point Chern number for different lattice sizes ($L=42, 52, 72)$, while the PBC/OBC Chern marker maps in the right-hand panel sample, respectively, the topological phase with few vacancies ($\rho_v=0.02$), the region around the transition ($\rho_v=0.2$), and the trivial phase at high vacancy density ($\rho_v=0.38$). As for Anderson disorder, also in the case of vacancies, the local topology is rather homogeneous, except around the phase transition that is characterized by patches and finite-size effects.

%% file: sm_text.tex
\renewcommand{\figurename}{Supplementary Fig.}
\renewcommand{\tablename}{Supplementary Tab.}
\renewcommand\thefigure{\arabic{figure}}
\renewcommand\thetable{\arabic{table}}

\renewcommand\thesubsection{Supplementary Note \arabic{subsection}}

\renewcommand\theequation{S\arabic{equation}}

\setcounter{equation}{0}
\setcounter{figure}{0}
\setcounter{table}{0}
\setcounter{subsection}{0}
\newcommand\kp{{\bm k}_\parallel}

\section{Fluctuations and correlations in quantum spin Hall insulators}
Here, we consider the Kane-Mele (KM) model~\cite{kane_quantum_2005,kane_z2_2005} on the honeycomb lattice in the presence of Anderson disorder~\cite{abrahams_prl_1979}. For the pristine case, the real nearest-neighbor hopping is set to $t=1$, the diagonal Kane-Mele spin-orbit coupling (SOC) is a complex next-nearest neighbor hopping  with a spin-dependent amplitude $\lambda_{SOC} = 0.05$ and proportional to the Pauli matrix $\sigma_{z}$. The staggered on-site potential $\Delta$ is such that $\Delta / \lambda_{SOC} = 3$, while the Rashba SOC term, a complex nearest-neighbor hopping with off-diagonal spin components, is set to $\lambda_{R} / \lambda_{SOC} = 1$. For this set of parameters, the system is in the topological phase. Disorder is introduced through a random on-site term with uniform distribution $w_i \in \left[ -\frac{W}{2},\frac{W}{2} \right]$, where $W$ is the disorder strength.

Quantum spin Hall insulators are characterized by a $\mathbb Z_2$ invariant protected by time reversal symmetry, here we study the topology of the system globally via the single-point spin-Chern number~\cite{Favata_2023}  $\nu_{sp}$, and locally by employing the local spin-Chern marker~\cite{z2_marker} $\nu(\mathbf r)$. Both methods are based on the spin-Chern number, introduced in Ref.~\cite{prodan_NJP_2010}, that is well-defined if a spectral gap in the projected spin operator $\mathcal PS_z\mathcal P$ exists. If that is the case, the subspace spanned by the eigenvectors of $\mathcal PS_z\mathcal P$ with positive (negative) eigenvalues is characterized by an integer-valued individual Chern number $C_+$ ($C_-$). These can then be promoted to individual local Chern markers using the methods introduced in Ref.~\cite{z2_marker} (and summarized in the main text), both for open (OBC) and periodic (PBC) boundary conditions, allowing us to define a local $\mathbb Z_2$ invariant in terms of a spin-Chern marker as
\begin{eqnarray}\label{eq:lscm}
  \nu(\mathbf r)=\frac{C_+(\mathbf r)-C_-(\mathbf r)}{2}\mod 2.
\end{eqnarray}
The spectral gap in $\mathcal PS_z\mathcal P$ is typically non-vanishing also when $S_z$ is not a good quantum number (e.g., in the presence of Rashba SOC), but in general the existence of a topological phase does not imply the existence of such a gap. Although, in principle, the gap of $\mathcal PS_z\mathcal P$ could also close in an insulating phase, we check that this does not occur in any of the simulations performed here. Additionally, the spin-Chern marker converges faster with the system size and it is more practical than a $\mathbb Z_2$ marker solely based on time reversal symmetry; both have been introduced and compared in Ref.~\cite{z2_marker}. For these reasons, we choose the spin-Chern number and marker to study the disorder-driven phase transitions in the model. However, our methods and conclusions are more general and would equally hold for the $\mathbb Z_2$ marker based on time reversal symmetry, which may be useful in scenarios where the spin-Chern marker is ill-defined.

\begin{figure*}[t]
  \includegraphics[width=\textwidth]{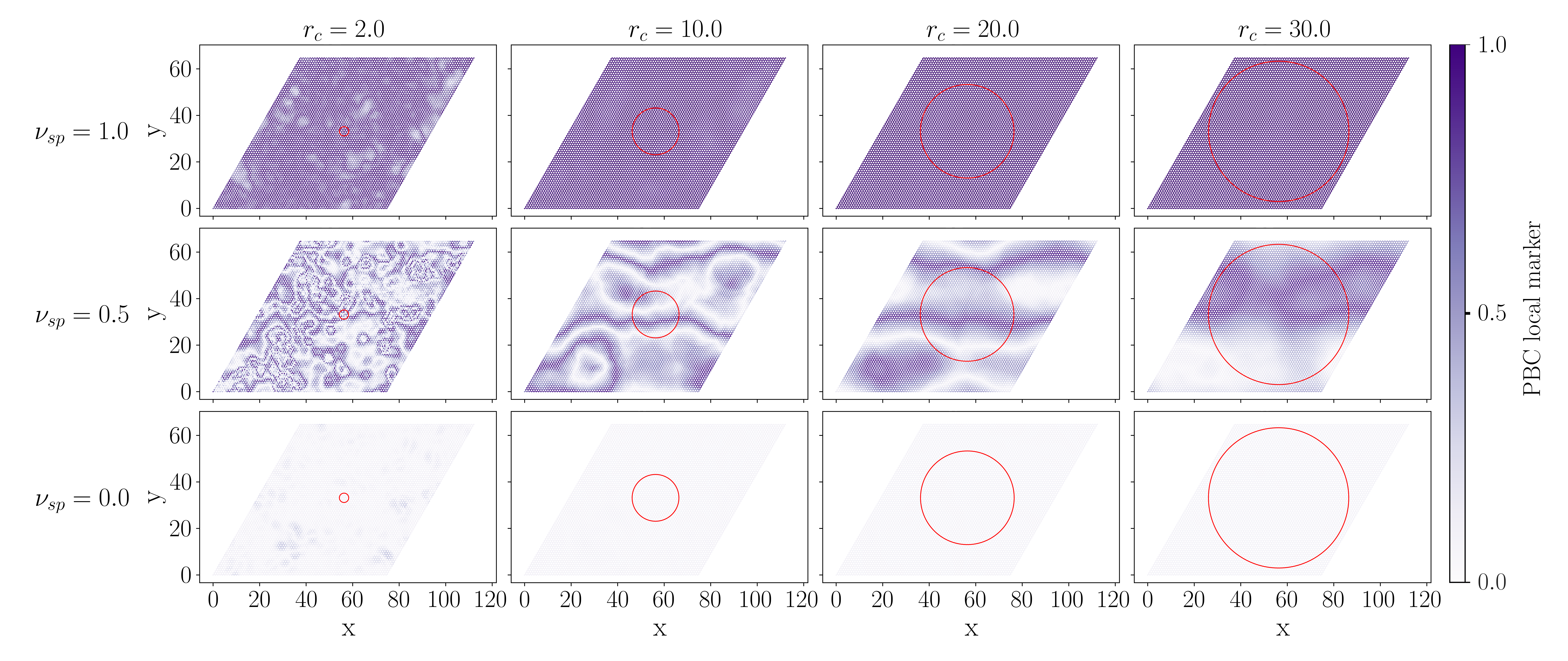}
  \caption{\label{sfig:fluc_corr_km}Macroscopic average of the local $\mathbb{Z}_2$ invariant in quantum spin Hall insulators: Real-space maps for the spin-Chern marker in PBC are shown for three different values of the disorder strength $W/t$ that is added to the Kane-Mele model (linear size $L=75$), corresponding to three different values of the single-point spin Chern number ($\nu_{sp}=1.0$ for $W/t = 3$, $\nu_{sp}=0.5$ for $W/t = 4$ and $\nu_{sp}=0.0$ for $W/t = 6.8$). The maps are drawn for increasing values of the radial cutoff $r_c$ within which the macroscopic average is performed. The cutoff $r_c$ is marked by the radius of the red circle centered in the middle of the sample. Short-scale fluctuations of the topological marker average out in the topological and trivial phase, while large-scale fluctuations remain at the transition.}
\end{figure*}

The KM model undergoes disorder-driven topological phase transitions, which are analogous to those observed in the Haldane model; these are discussed in terms of the single-point spin Chern number in Ref.~\cite{Favata_2023}. Supplementary Fig.~\ref{sfig:fluc_corr_km} displays real-space maps for the $\mathbb{Z}_2$ invariant $\nu(\mathbf r)$ in PBCs computed for increasing cutoff radii (red circles) ranging from $r_c=2$ to $r_c=30$, in the topological  ($\nu_{sp}=1.0$, $W/t = 3 $) and trivial ($\nu_{sp}=0.0$, $W/t = 4$) phase, as well as close to the topological phase transition ($\nu_{sp}=0.6$, $W/t = 6.8$). In constructing these maps, a macroscopic average is first taken over a cutoff radius 
$r_c$, followed by a modulo 2 operation applied to each marker on the map as dictated by the definition of the $\mathbb{Z}_2$ invariant [Eq.~\ref{eq:lscm}]. Finally, the results are mapped into the range $[0,1]$ to aid in visualization.
Fluctuations become very strong close to the phase transition, where finite-size effects are more visible. By averaging over larger regions, essentially coarse graining the topological markers, short-scale fluctuations  disappear both in the topological and trivial phase: already at $r_c=10$ these disordered systems appear topologically uniform. On the contrary, sufficiently close to the phase transition we observe large-scale fluctuations that persist even if the macroscopic average is performed over rather large regions (e.g., $r_c=30$).

\section{Correlation functions, exponents and finite-size effects}
In the main text, we compare the power-law decay of the correlation function $G(r)$ of local topological markers for different phase transitions. As part of the analysis, here we comment on the role of finite-size effects, which are known to become significantly pronounced near critical points. In particular, the long-range behavior of $G(r)$ is affected even for large\textemdash but finite\textemdash supercells: this is shown in the Supplementary Fig.~\ref{sfig:finite_size} for the topological-to-trivial transition in the Haldane model, while we observe similar behavior for other transitions. Specifically, correlations at large distances become slightly negative and approach a finite value, denoted by $G(\infty)$. This behavior is due to the very long correlation lengths close to the critical point, which become comparable with the simulation cell such that spurious correlations between periodic images are present. Crucially, in the thermodynamic limit, $G(\infty)$ goes to zero with the system size and $G(r)$ becomes positive definite everywhere. This is shown in the inset of Supplementary Fig.~\ref{sfig:finite_size}, where we plot the value of $G(\infty)$ as obtained by fitting the correlation function at the transition point with
\begin{eqnarray}\label{eq:g_r}
G (r) = G(\infty) + \frac{c}{r^p},
\end{eqnarray}
for increasing sample sizes, where the value of $c$ is also obtained by fitting. 
Hence, in Fig.~\ref{fig:critical} of the main text we subtract $G(\infty)$ from $G(r)$ when comparing the power-law decays, in order to visualize more clearly whether different transitions share the same exponent $p$ (i.e., the slope in a log-log plot) or not. Finally, Supplementary Fig.~\ref{sfig:finite_size} shows that the exponent $p$ extracted from the data does not strongly depend on the system size and converges to the value of $1/2$ for large simulation cells.

\begin{figure*}
  \includegraphics[width=0.5\linewidth]{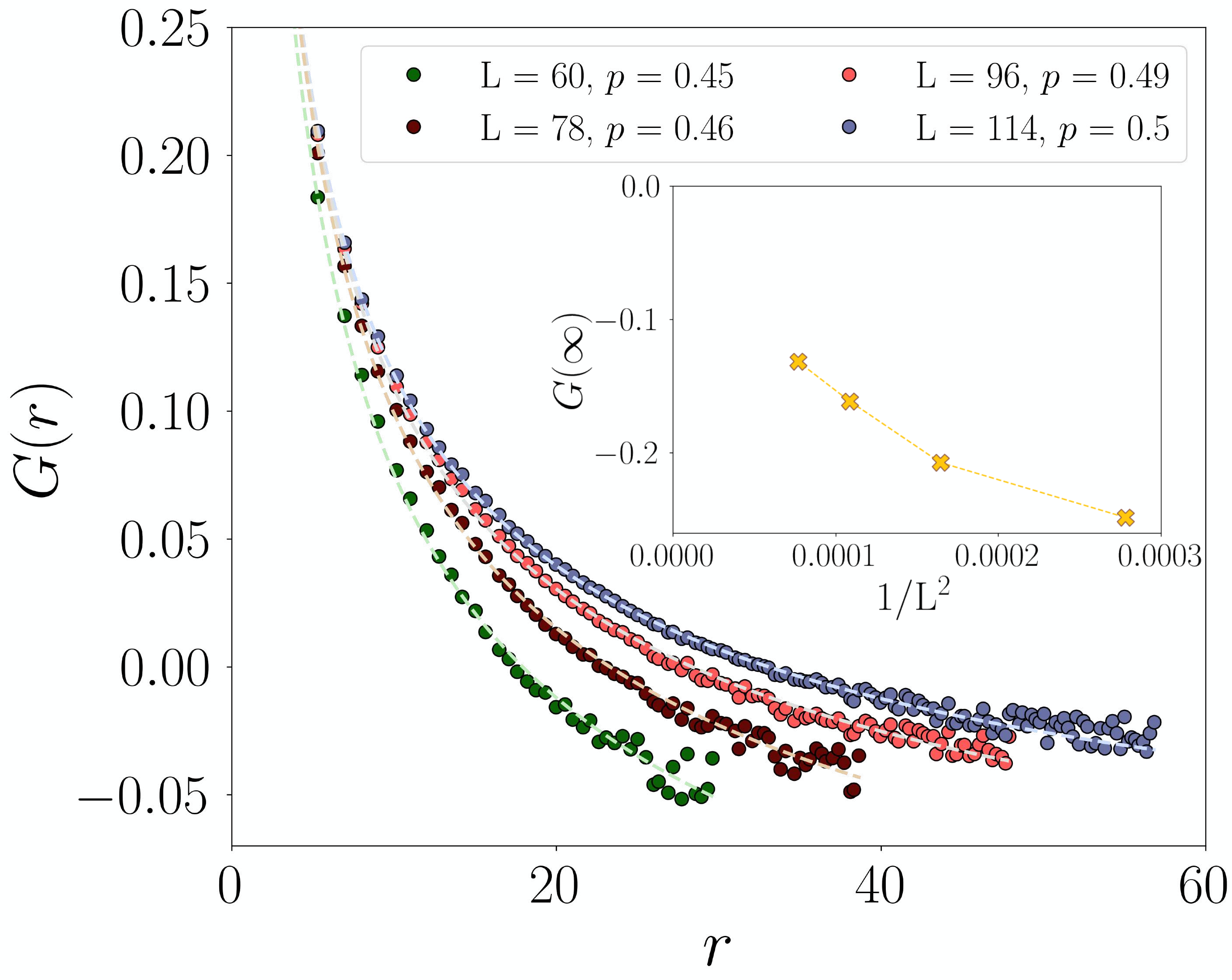}
  \caption{\label{sfig:finite_size} Correlation function $G(r)$ for different cell sizes $L$ in the Haldane model ($\Delta / t = 1.25$, $\phi= -\frac{\pi}{2}$) at the disorder-driven transition with $W / t = 6.5$. Dashed lines are obtained by power-law fitting with Eq.~\ref{eq:g_r}, from which a critical exponent $p \simeq \frac{1}{2}$ is consistently obtained for all sizes. The fitting also yields $G(\infty)$, which is shown in the inset to vanish in the thermodynamic limit.}
\end{figure*}